\documentclass{article}

\usepackage[english]{babel}
\usepackage{authblk}
\usepackage[letterpaper,top=2cm,bottom=2cm,left=3cm,right=3cm,marginparwidth=1.75cm]{geometry}
\usepackage{adjustbox}
\usepackage{amsmath}
\usepackage{graphicx}
\usepackage[colorlinks=true, allcolors=blue]{hyperref}
\usepackage{wasysym}
\usepackage{subfig}
\usepackage{soul}
\usepackage{floatrow}
\newfloatcommand{capbtabbox}{table}[][\FBwidth]

\usepackage[numbers,sort&compress]{natbib}

\title{High angular sensitivity X-ray phase-contrast microtomography of soft-tissue through a two-directional beam-tracking synchrotron set-up}
\author[1]{Carlos Navarrete-Le\'{o}n}
\author[2]{P. Stephen Patrick}
\author[1]{Adam Doherty}
\author[1]{Harry Allan}
\author[1]{Silvia Cipiccia}
\author[3]{Shashidhara Marathe}
\author[3]{Kaz Wanelik}
\author[1]{Michela Esposito}
\author[1]{Charlotte K. Hagen}
\author[1]{Alessandro Olivo}
\author[1]{Marco Endrizzi}

\affil[1]{Department of Medical Physics and Biomedical Engineering. University College London, Malet Place, Gower Street, London, WC1E 6BT, United Kingdom}
\affil[2]{Centre for Advanced Biomedical Imaging, Division of Medicine, University College London, Paul O'Gorman Building, 72 Huntley Street, WC1E 6DD, United Kingdom}
\affil[3]{Diamond Light Source, Harwell Science and Innovation Campus, Fermi Avenue, Didcot, OX11 0DE, United Kingdom}

\date{}

\begin{document}
\maketitle

\begin{abstract}
Two-directional beam-tracking (2DBT) is a method for phase-contrast imaging and tomography that uses an intensity modulator to structure the X-ray beam into an array of independent circular beamlets that are resolved by a high-resolution detector. It features isotropic spatial resolution, provides two-dimensional phase sensitivity, and enables the three-dimensional reconstructions of the refractive index decrement, $\delta$, and the attenuation coefficient, $\mu$. In this work, we report on the angular sensitivity and the spatial resolution of 2DBT images in a synchrotron-based implementation. In its best configuration, we obtained angular sensitivities of $\sim$20 $n$rad and spatial resolution of at least 6.25 $\mu$m in phase-contrast images. We also demonstrate exemplar application to the three-dimensional imaging of soft tissue samples, including a mouse liver and a decellularised porcine dermis.
\end{abstract}

\section*{Introduction}

X-ray phase-contrast tomography (XPCT) is a non-destructive imaging technique that enables the 3D visualization of materials and tissues composed of low-Z elements. Through generating contrast also from the phase shift induced in the x-ray wavefront, this technique allows for the visualization of details otherwise undetectable by using a conventional, attenuation-based approach to generating image contrast \cite{Paganin2021}. Notable examples have been demonstrated in the biomedical field using synchrotron radiation, including brain \cite{Pinzer2012, Topperwien2020}, lung \cite{Xian2022, Reichmann_2023}, kidney \cite{Zdora2020, Walsh2021}, breast \cite{Zhao2012, Baran2018, AranaPena2023}, and oesophagus \cite{Hagen2015} imaging, amongst others.

Several methods have been developed at synchrotron radiation facilities for XPCT including crystal-based interferometric methods, propagation-based imaging methods, analyzer-based imaging methods, grating-based interferometric methods, speckle-based imaging methods, and non-interferometric mask-based methods \cite{bonse1965, Momose1996, Snigirev1995, Nugent1996, Chapman1997, Momose2003, Morgan2012, Berujon2012, Olivo2001, Morgan2011, Vittoria2015}.

Here we focus on two-directional beam-tracking (2DBT), which belongs to the category of non-interferometric mask-based methods. The method was first proposed by a patent in the mid-90s \cite{Wilkins_1995} and it shares similarities with the Shack-Hartmann wave-front sensor. It requires the use of a single optical element (modulator), which structures the beam into an array of independent beamlets that are then resolved by a high-resolution detector. The modulator is placed before the sample, contrary to other implementations \cite{BeganiProvinciali2020, BeganiProvinciali2021}, meaning that all photons reaching the sample contribute to image formation, limiting the absorbed dose.

In this approach, the radiation field intensity modulation, coupled with a dedicated data analysis methodology, provides a way to measure how the sample affects the intensity and position of each X-ray beamlet. A shift in their position is interpreted as a refraction effect, which is related to the first derivative of the phase shift imposed by the sample. A change in intensity is interpreted as attenuation of the X-ray beam. 

This approach provides aperture-driven \cite{Diemoz_2014, Hagen2018, Esposito2022} and isotropic \cite{NavarreteLeon2022} spatial resolution; and with dedicated mask designs alongside efficient acquisition schemes, scanning time can be improved for optimal acquisition \cite{Lioliou2022, Lioliou2023}, and fly-scan data acquisition schemes. We note that the approach allows also for X-ray dark-field imaging \cite{Dreier2020}. The method was recently demonstrated in a compact laboratory set-up \cite{Navarrete-Leon2023}, within a small ($<$1 m) footprint and by using a low power (10 W) source. Here we report on an implementation that made use of synchrotron radiation, and that we found suitable for high-sensitivity measurement of phase gradients, providing excellent contrast for the visualisation of the morphology in soft tissue samples. We report on the angular sensitivity of this approach, characterised as a function of exposure time and system geometry, as well as its spatial resolution, estimated with Fourier ring correlation on tomographic reconstructions. We demonstrate exemplary application to the three-dimensional imaging of soft tissue samples, including both a formalin-fixed sample of mouse liver, as well as a decellularised, iodine-stained porcine dermis. For both biological samples, phase-contrast imaging enhanced the visibility of key physiological features above attenuation-based images acquired with equivalent x-ray exposure.


\begin{figure*}[t]
\centering
\includegraphics[width=0.7\textwidth]{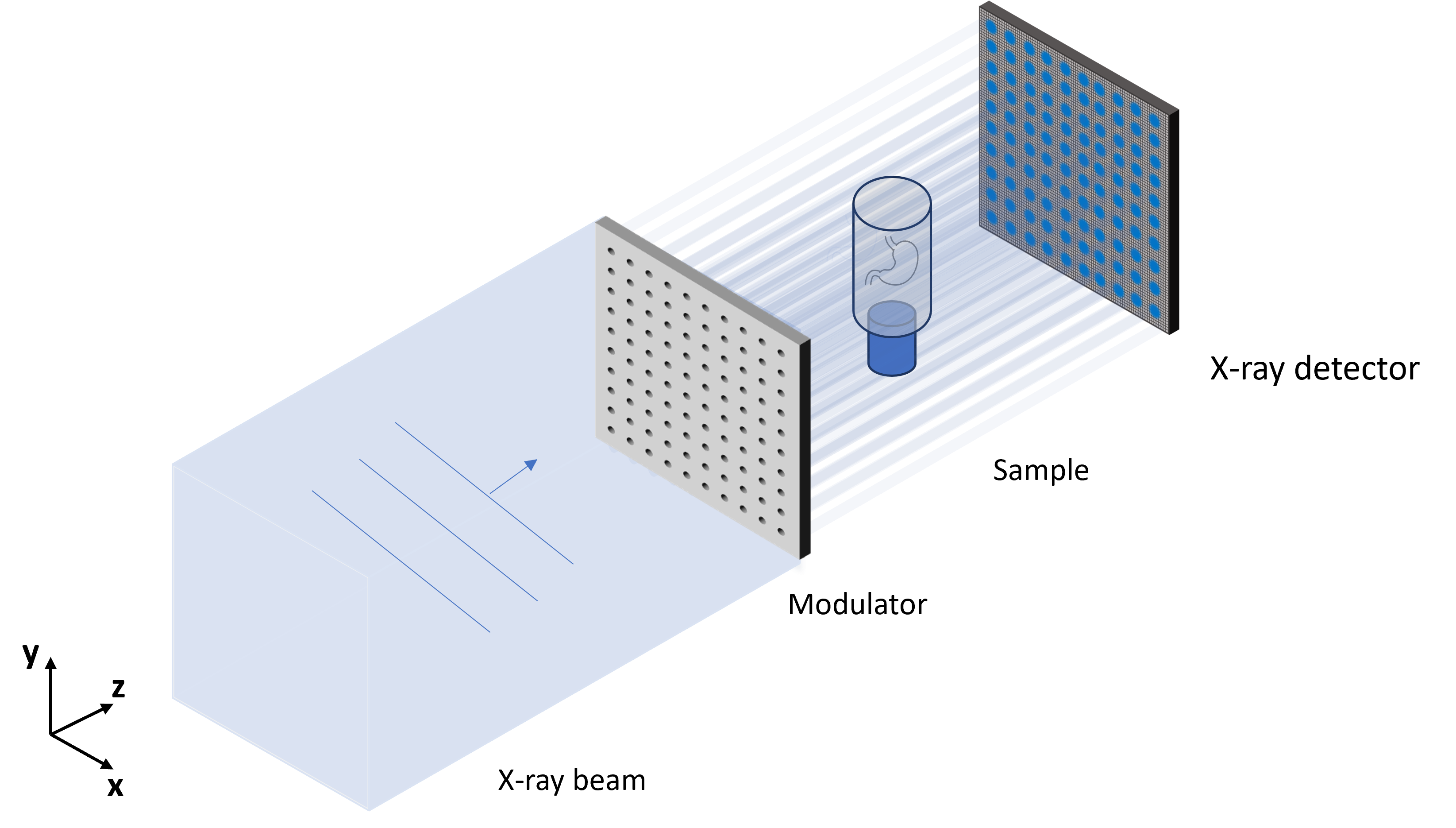}
\caption{Schematic diagram of the two-directional beam-tracking experimental set-up.}
\label{fig1:set-up}
\end{figure*}

\section*{Materials and methods}

\subsection*{2DBT x-ray set-up}
The XPCT set-up is presented in Figure \ref{fig1:set-up}. The experiments were carried out at Diamond Light Source Beamline I13-2. The angular sensitivity measurements were done with a mean energy of 16 keV from a filtered pink beam with a Silicon mirror and filters of 1.34 mm Pyrolytic graphite, 1.4 mm Aluminium, and 0.042 mm Niobium. For the biological specimens, the mean energy was increased to 27 keV to reduce sample damage by changing to a Platinum mirror and filters of 1.34 mm Pyrolytic graphite and 3.2 mm Aluminium.

The sample was placed roughly 221 m from the source, and the modulator was placed 15 cm upstream of the sample. The modulator is fabricated with laser-ablation from a 100 $\mu$m thick tungsten foil (Goodfellow), and has a period of 50 $\mu$m. The apertures have a conical shape with diameters of 15 $\mu$m in the front and 30 $\mu$m in the back. The detector is a pco.edge 5.5 camera coupled to a scintillator-objective combination with an effective pixel size of $2.6\times2.6$ $\mu$m$^2$ and a field of view of $6.6\times5.6$ cm$^2$.

\subsection*{Angular sensitivity measurements}

The angular sensitivity was assessed with a custom-built phantom composed of soda-lime glass microspheres of 50 $\mu$m diameter (Fischer Scientific, monodisperse) embedded in wax and polyethylene foam. To study the sensitivity as a function of the object-to-detector distance ($z_{od}$), the sample was imaged at the following distances: $z_{od}$=\{2.5, 7.5, 17.5, 37.5, 77.5\} cm, by moving the detector. The sample was moved in a $10\times10$ grid with steps of 5 $\mu m$ in the $xy$ plane. At each sample position, $10\times0.1$ s frames were acquired, and flat and dark images were taken before and after each scan. The frames were used to assess sensitivity as a function of the exposure time. The assessment of the angular sensitivity during imaging was carried out by calculating the mean and standard error of the standard deviation of the measured refraction angles in an area without the sample, for which eight different windows of $5\times40$ pixels were used.

\subsection*{Tomography of unstained and stained ex-vivo tissues}
Two biological samples were imaged: a mouse liver and a hernia mesh, consisting of decellularised porcine dermis. The liver was fixed in 4\% para-formaldehyde for 24 hours upon dissection from a 2 month old C57BL/6 mouse (Charles River Laboratories), then stored in 0.9\% saline. The decellularised porcine dermis (Xenmatrix$^{TM}$, Bard), was stained in 3\% Lugol's iodine solution (Scientific Laboratory Supplies) in phosphate buffered saline (Gibco) for 24 hours before storage in 0.9\% saline. Both liver and decellularised dermis were prepared for imaging by embedding in 1\% agar (Thermo Scientific Chemicals). The samples were between 3.1 and 3.8 mm wide and they were scanned by acquiring 1200 projections while rotating over 180° in a fly-scan fashion with an exposure time of 0.15 s per projection. This was repeated at different modulator sub-pitch displacements to increase sampling. The modulator was raster-scanned in $8\times8$ positions, by using 6.25 $\mu$m displacements both in $x$ and $y$. This led to a total exposure time of $1200\times8\times8\times0.15$ s = 3.2 h for each sample. Flat and dark images were acquired at each modulator position, before and after rotating the sample. Note that this is different from radiography, in which it was the sample that was moved. The detector was placed 128 cm away from the sample to further increase the angular sensitivity, which was also assessed for this configuration with eight different windows of $8\times8$ pixels.

\subsection*{Data Analysis}
The transmission, refraction in $x$, and refraction in $y$ images were obtained by selecting a window of $20\times20$ pixels around each beamlet and comparing the intensities with ($I_s(x,y)$) and without ($I_0(x,y)$) the sample in the beamlet. The transmission was calculated by dividing the sum of the intensities in the windows: $t=\sum_{xy} I_s(x,y)/\sum_{xy} I_0(x,y)$, and the two refraction images by measuring the displacements ($\Delta_x$, $\Delta_y$) between the beamlets with a subpixel cross-correlation algorithm \cite{GuizarSicairos2008}. This was performed for all images acquired at each sample or modulator position, which were then stitched to obtain an image with higher sampling \cite{Navarrete-Leon2023}.

Assuming small refraction angles and under a geometrical optics approximation, the refraction angle, $\alpha_{xy}$, is related to the displacements $\Delta_x$ and $\Delta_y$ and the orthogonal gradients of the phase shift, $\Delta \Phi_{x,y}$, by:

\begin{equation}
    \alpha_{x,y}=\frac{\Delta_{x,y}}{z_{od}}=\frac{\Delta \Phi_{x,y}}{k},
\end{equation}
where $z_{od}$ is the object-to-detector distance and $k$ is the wavenumber. This allows us to obtain the phase shift $\Delta \Phi$ by integrating the two gradients through a Fourier space method \cite{Kottler2007}). 

The retrieved quantities $t$ and $\Delta \Phi$ are linked to integrals along the photon path of the linear attenuation coefficient ($\mu$) and the real part of the refractive index ($\delta$) in the following way:
\begin{equation}
    \label{eq:mu}
    -\ln t(x,y) = \int_o\mu(x',y',z') dz
\end{equation}

\begin{equation}
    \label{eq:delta}
    -\frac{\Delta \Phi (x,y)}{k} = \int_o\delta(x',y',z')dz.
\end{equation}
Therefore, for the biological specimens, volumes of $\mu$ and $\delta$ were obtained from the projections taken at different viewing angles using the filtered back projection (FBP) implementation of the Astra toolbox \cite{astra_2015}.

For both the $\mu$ and $\delta$ volumes, the spatial resolution of slices in the three orthogonal planes was estimated using an implementation of Fourier ring correlation (FRC) \cite{nieuwenhuizen2013measuring} provided as part of the BIOP ImageJ plugin \cite{HerbFRC}. Independent inputs were provided to the algorithm by reconstruction of two volumes, each using half of the available projections. To reduce the noise of the FRC estimate, the resultant curves of 5 adjacent representative slices from the middle of the volume were averaged. Resolutions are stated using the 3-$\sigma$ criterion, expressing the spatial frequency at which the FRC curve exceeds by 3 standard deviations the expected correlations within the random background noise \cite{van2005fourier}.

\begin{figure*}[h]
\centering
\includegraphics[width=\textwidth]{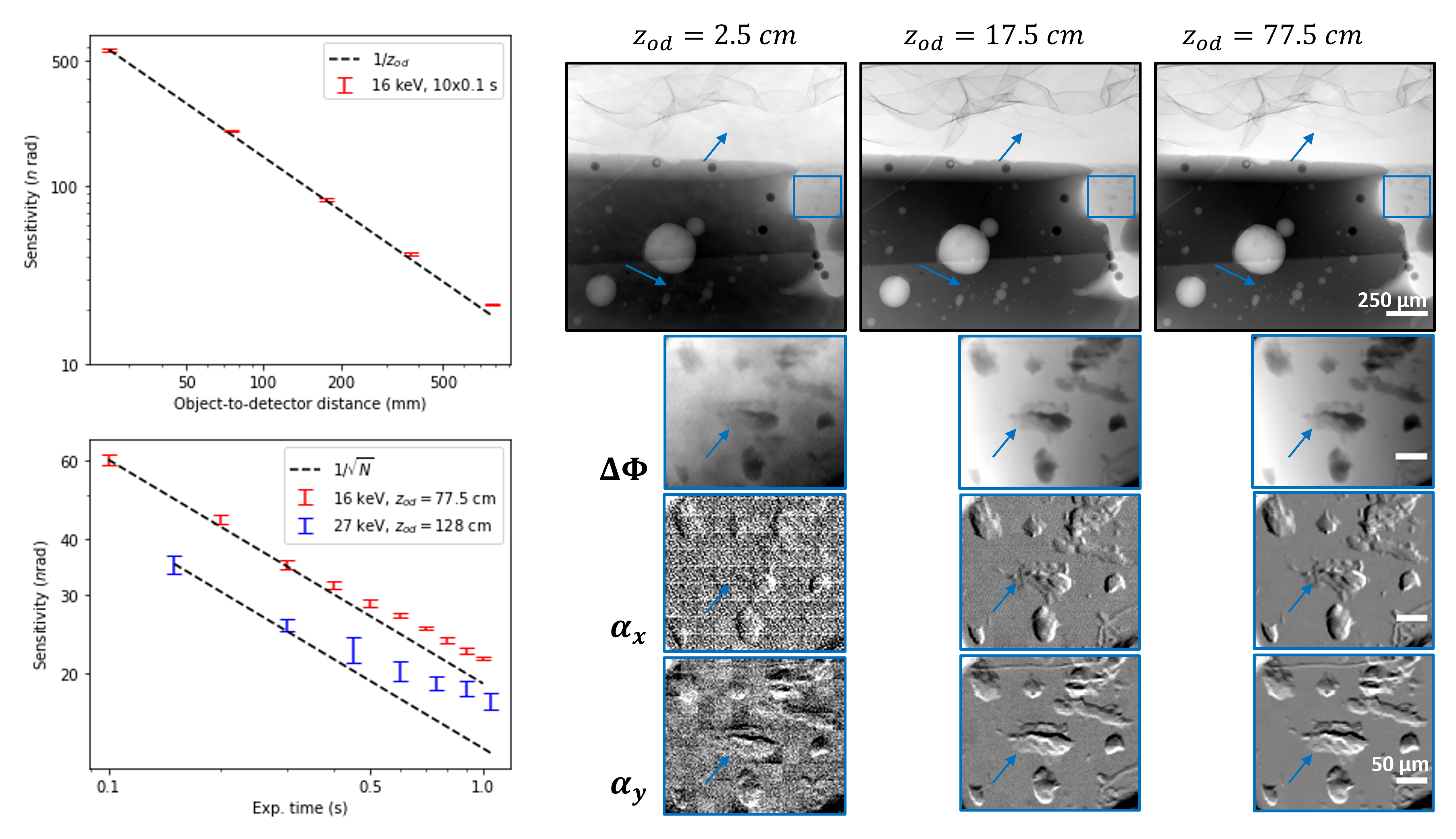}
\caption{Angular sensitivity of the method as a function of (a) object-to-detector distance and (b) exposure time for different system configurations. (c) Phase images of the phantom (polyethylene foam, and microspheres and air bubbles embedded in wax) are shown for increasing object-to-detector distances. The improvement in angular sensitivity reveals interfaces in the foam and small bubbles in the wax substrate, as pointed out by the arrows. d) An inset in the phase image is shown, along with the two refraction images, for further demonstration of thin wax deposits on the substrate being unveiled with increasing sensitivity.}
\label{fig2:sensitivity}
\end{figure*}

\section*{Results and discussion}
\subsection*{Angular sensitivity}
The results from the angular sensitivity measurements are reported in Figure \ref{fig2:sensitivity}. The sensitivity is shown for different object-to-detector distances (Fig. \ref{fig2:sensitivity}a) and for an increasing number of integrated frames for both energy configurations (Fig \ref{fig2:sensitivity}b). We observe that the smallest resolvable angle decreases proportionally with increasing propagation distance ($\propto 1/z_{od}$) within the range of propagation distances explored, as expected from the geometrical optics approximation used in Equation 1. We note a small deviation from this trend at 77.5 cm of propagation distance and we observe it is associated with a small decrease in visibility, from 85\% to 82\%. Longer propagation distances offer a further increase in the angular sensitivity, however, the angular sensitivity is expected to increase at a relatively slower rate beyond this point. In addition, we also observed that when integrating a few frames, the sensitivity goes inversely with the square root of the number of counts ($\sqrt{N}$). This is the expected trend from Poisson statistics, indicating that this is the dominant noise source up to accumulations of 400 ms; beyond this point, additional noise sources become significant in limiting the smallest measurable refraction angle. The smallest angle with a mean energy of 16 keV was measured with a combination of $z_{od}=77.5$ cm and $10\times0.1$s frames, for which we report an angular sensitivity of $21.6\pm0.2$ $n$rad. In the conditions for tomographic imaging at 27 keV, the angular sensitivity benefits from the increased propagation distance of $z_{od}=128$ cm, and we measured $35\pm2$ $n$rad with only $1\times0.15$s frame.

The effect of increasing angular sensitivity on image quality can be observed in Figures \ref{fig2:sensitivity}c,d, where the integrated phase images are presented along with both refraction images of the smaller, highlighted region of interest. The increasing angular sensitivity unveils various interfaces in the foam and small bubbles in the wax substrate, as pointed out by the arrows in Fig. \ref{fig2:sensitivity}c. This is further demonstrated with the insets in Fig. \ref{fig2:sensitivity}d, where the two refraction images show increasingly lower noise levels as propagation distance is increased, which in this case reveals thinner deposits of wax on the substrate as angular sensitivity increases.

\begin{figure*}[h]
\centering
\includegraphics[width=0.9\textwidth]{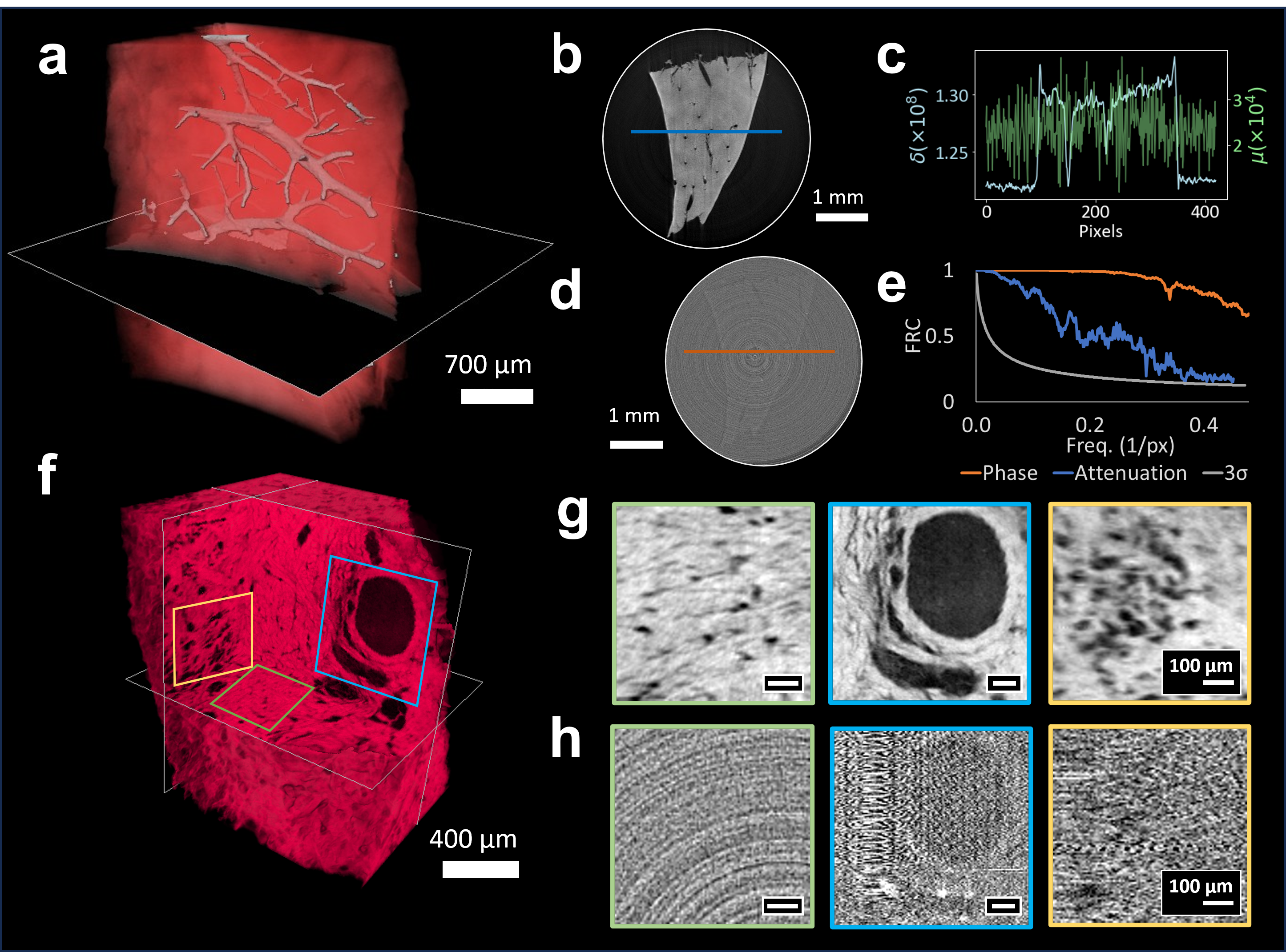}
\caption{X-ray phase-contrast and attenuation tomography of biological soft-tissues shown as 3D-rendered volumes and in representative perpendicular planes. (a, b) phase-contrast, and (d) attenuation tomography of mouse liver. (c) Line profile of signal through indicated liver cross sections of b and d showing improved contrast to noise ratio in the phase contrast. (e) Fourier ring correlation curve calculated from two independent reconstructions of (b) and (d) showing an increased resolution for phase-contrast. (f) 3D render of phase contrast tomography of decellularised porcine dermis, showing perpendicular cross-sections of  (g) phase-contrast and (h) attenuation.}
\label{fig3:tomography}
\end{figure*}

\subsection*{Tomography of biological soft tissues}

The attenuation and phase contrast tomographic reconstructions of the liver tissue and decellularised dermis samples are presented in Figure \ref{fig3:tomography}. The significant increase in contrast-to-noise ratio achieved by means of phase contrast is evident across all slices and for both stained and unstained samples. While the image quality enhancement from phase contrast was particularly evident in the unstained (i.e. poorly absorbing) liver tissue, even tissue optimised for attenuation-based imaging using an iodine stain still showed notable improvement with phase contrast.  In terms of physiological features identifiable in the liver sample, phase-contrast gave clear visualisation of the hepatic portal vein, arteries, and bile ducts (Fig \ref{fig3:tomography}a,b), while these were not discernible on the attenuation-based images (Fig. \ref{fig3:tomography}c). For the stained sample of the decellularised dermis, while the lumen of the hair follicles was discernible in both phase (Fig. \ref{fig3:tomography}f,g) and attenuation-based images (Fig. \ref{fig3:tomography}h), phase-contrast improved the clarity of the structure of the hair follicles including the layers of the root sheath, sebaceous gland, and fibre alignment around the dermic sheath (Fig \ref{fig3:tomography}f-h).  

Figure \ref{fig3:tomography}e displays the FRC curves obtained from the unstained liver tissue axial slices, as illustrated in Figure \ref{fig3:tomography}b and Figure \ref{fig3:tomography}d. The attenuation curve intersects the threshold at 0.4 px$^{-1}$, indicating a spatial resolution of 16 $\mu$m. Meanwhile, the phase-contrast curve fails to intersect with the threshold, suggesting that in this case the spatial resolution is sampling limited and is at least equal to the sampling pixel size of 6.25 $\mu$m. We note that the FRC curves in the orthogonal cross-sections showed comparable trends. We interpret the disparity between the FRC curves obtained through the attenuation- and phase-contrast tomography as a consequence of the noise and contrast dependence inherent in the FRC resolution metric. High-spatial frequency image features are unable to surpass the noise threshold in the noisier, low-contrast attenuation volume, whereas the much higher signal-to-noise ratio achieved with phase contrast allows for separating even the smallest features from the background. We also note that by splitting the projection dataset to obtain independent volumes, the angular tomographic sampling has also been halved to 600 projections. Although the full dataset was largely oversampled in terms of viewing angles, the halved dataset is undersampled with respect to the Nyquist sampling theorem for samples between 500 and 600 pixels of width. As such, the FRC result should still be considered a conservative estimate of the achievable resolution.

\section*{Conclusion}
We have here studied the angular sensitivity and spatial resolution in images obtained through a 2DBT synchrotron set-up and shown the potential of the method for volumetric imaging of soft tissues with poor attenuation contrast. We report angular sensitivities of $\sim$20 $n$rad at 1s exposure time, 77.5 cm of propagation distance, and 16 keV mean energy; and $\sim$35 $n$rad at 150 ms, 128cm, and 27 keV mean energy. Our results indicated that the geometrical-optics approximation used by the phase-retrieval algorithms is well satisfied within 1 m of propagation distance. We have also shown sub-aperture spatial resolution in phase-contrast tomography, which was observed to be limited by sampling to a factor 2.4x better than the apertures in the modulator. We note that this indicates a spatial resolution improvement with respect to what was previously modelled \cite{Diemoz_2014}. These results provide a basis for future experimental designs with the 2DBT method, especially for identifying optimal trade-offs between angular sensitivity, spatial sampling, and acquisition time. They also provide a basis for comparison with similar imaging methods. 


\section*{Acknowledgements}
We gratefully acknowledge Diamond Light Source for time on Beamline I-13 under Proposal MG30748. This work was supported by the EPSRC EP/T005408/1; Wellcome Trust 221367/Z/20/Z; the National Research Facility for Lab x-ray CT (NXCT) through EPSRC grants EP/T02593X/1 and EP/V035932/1. PSP acknowledges funding by MRC grant MR/R026416/1, EPSRC Knowledge Exchange and Innovation Fund award EP/R511638/1, and Wellcome Trust/UCL Devices \& Diagnostics Therapeutic Innovation Network Pilot Data fund (562448 – Linked to Lead 553191). CKH was supported by the Royal Academy of Engineering under the Research Fellowship scheme. AO is supported by the Royal Academy of Engineering under their Chair in Emerging Technologies scheme (CiET1819/2/78). 

\bibliographystyle{ieeetr}
\bibliography{references}
\end{document}